# Topological nature of the Kondo insulator SmB₆ and its sensitiveness to Sm vacancy

W. K. Park[1,*], J. A. Sittler,[1,2,a] L. H. Greene,[1,2] W. T. Fuhrman,[3] J. R. Chamorro,[3,4] S. M. Koohpayeh,[3] W. A. Phelan,[3,b] T. M. McQueen[3,4,5]

[1]*National High Magnetic Field Laboratory, Florida State University, Tallahassee, FL 32310, USA*

[2]*Department of Physics, Florida State University, Tallahassee, FL 32306, USA*

[3]*Institute for Quantum Matter, Department of Physics and Astronomy, Johns Hopkins University, Baltimore, MD 21218, USA*

[4]*Department of Chemistry, Johns Hopkins University, Baltimore, MD 21218, USA*

[5]*Department of Materials Science and Engineering, Johns Hopkins University, Baltimore, MD 21218, USA*

The true topological nature of the Kondo insulator SmB₆ remains to be unveiled. Our previous tunneling study not only found evidence for the existence of surface Dirac fermions, but it also uncovered that they inherently interact with the spin excitons, collective excitations in the bulk. We have extended such a spectroscopic investigation into crystals containing a Sm deficiency. The bulk hybridization gap is found to be insensitive to the deficiency up to 1% studied here, but the surface states in Sm-deficient crystals exhibit quite different temperature evolutions from those in stoichiometric ones. We attribute this to the topological surface states remaining incoherent down to the lowest measurement temperature due to their continued interaction with the spin excitons that remain uncondensed. This result shows that the detailed topological nature of SmB₆ could vary drastically in the presence of disorder in the lattice. This sensitiveness to disorder is seemingly contradictory to the celebrated topological protection, but it can be understood as being due to the intimate interplay between strong correlations and topological effects.

[*]Author to whom all correspondence should be addressed: wkpark@magnet.fsu.edu

[a]Present address: Department of Physics, University of California Santa Cruz, Santa Cruz, CA 95064, USA

[b]Present address: Los Alamos National Laboratory, Los Alamos, Mail Stop E574, Los Alamos, NM 87545, USA

# I. INTRODUCTION

Topological materials defy description by the conventional Landau-Ginzburg paradigm based on broken symmetries [1, 2]. Instead, they can be classified by a completely new scheme, namely, topological invariants characterizing their band structure. A topological insulator (TI) is insulating in the bulk, but its surface is a robust conductor due to the topological surface states (TSSs) protected by spin-momentum locking [3]. In an *f*-orbital-based Kondo insulator (KI) [4], strong correlations play a key role in reducing the hybridization gap (HG) by bringing the hybridized band edges close to each other, enhancing the effective mass [5, 6]. Then, combined with the inherently large spin-orbit coupling, band inversions can occur at high-symmetry points, possibly transforming the KI into a topological phase, namely TKI [7]. The true nature of TKIs, a representative group of correlated topological phases, remains to be unveiled, posing a big challenge [6, 8], in sharp contrast to their weakly correlated counterparts, much of whose underlying physics is now well understood [2].

$SmB_6$ has been a focus of intensive study recently as a prime candidate for a TKI [6, 8]. Although the low-temperature resistance plateau is now firmly established as due to the TSSs, the exact topological nature is not known yet with the origin of quantum oscillations remaining largely controversial [9-12]. This is due to multi-faceted challenges, including (i) materials science issues in the bulk [13, 14] and/or at the surface (e.g., polar nature) [15, 16], and (ii) the intriguing interplay between strong correlations and topology that is not yet well understood. Consequently, despite the early observation of a HG in several spectroscopic studies, including angle-resolved photoemission spectroscopy (ARPES) [17-23], scanning tunneling spectroscopy (STS) [15, 24, 25], and point-contact spectroscopy (PCS) [26], spectroscopic signatures for the TSSs remained obscure, as discussed in the supplemental material (SM) [27]. Our recent study employing planar tunneling spectroscopy (PTS) [28] has provided clear evidence for the surface Dirac fermions, in good agreement with band calculations [29-32] and a quantum oscillation study [9]. Remarkably, the TSSs were found inherently to interact with the spin excitons (SEs) [33], collective bulk excitations, in agreement with a theoretical [34] and an ARPES [35] studies.

One of the remaining crucial questions is how the topological nature of $SmB_6$, or TKIs in general, is influenced by a disorder in the (Kondo) lattice. This is a particularly important topic since Sm vacancies are known to occur easily in single crystals grown by the floating-zone (FZ) technique [13, 14]. Previously, impurities without *f*-electrons were speculated to induce in-gap states in the bulk, and thus to be responsible for the low-temperature metallicity [36, 37]. However, now that it is known to be due to the TSSs in stoichiometric crystals [6, 8], the role of Sm deficiency needs to be understood more comprehensively by taking the topological aspect into account. More specifically, studying how the SEs evolve with increasing Sm deficiency and consequently how their effect on the TSSs changes may help elucidate the interplay between strong correlations and topology. In this paper, we report a PTS study on Sm-deficient $SmB_6$ crystals and we compare it with our previous results from stoichiometric crystals [28, 38]. While the bulk HG largely remains the same, the TSSs are found to undergo distinct temperature evolutions due to drastic changes in their interaction with the SEs.

## II. RESULTS

The normalized dc electrical resistance, $R_n(T) \equiv R(T)/R(300K)$, measured across the polished (001) surface is plotted in Fig. 1 for three $SmB_6$ crystals that are also used in our PTS study. One is a stoichiometric flux-grown (FG) crystal studied previously [28, 38]. The other two are grown by the floating-zone technique, labeled by the nominal purity of the starting growth material as FZ-Pure and FZ-Def (FZs collectively), respectively. The FZs are known to have Sm vacancies up to 1% [13, 14] regardless of the nominal purity, which can be inferred from the similar behaviors they exhibit throughout this study. At high temperature, the three crystals show an almost identically insulating behavior including a broad hump at 15 – 20 K as marked by the arrow, below which a noticeable slope change occurs (see the inset). The activation-type energy gaps (in meV) extracted from an Arrhenius analysis for the temperature range higher ($\Delta_h$) and lower ($\Delta_l$) than the hump temperature are ($\Delta_h$, $\Delta_l$) = (5.6, 3.1), (5.4, 2.9), and (5.2, 3.3) for the FG, FZ-Pure, and FZ-Def, respectively. Note that below ~10 K, the $R_n(T)$ curves become distinct: While the FG forms a pronounced plateau below ~4 K after a rapid increase by five orders of magnitude [39], the FZs show an increase only by two to three orders of magnitude followed by a mere *slowdown* below ~7 K. This discrepancy is likely to be associated with the Sm deficiency [13, 14], but transport studies alone cannot address whether the slowdown in the FZs is still due to the TSSs or the in-gap states.

For spectroscopic investigations into this question, tunnel junctions are prepared by following the same procedure as before with Pb as the counter-electrode [28, 38]. It is noteworthy that some of the aforementioned surface issues were resolved by plasma-oxidizing the crystal to form a high-quality $B_2O_3$ tunnel barrier beneath which the TSSs reform [27]. The differential conductance, $g(V) \equiv dI/dV$, normalized by the parabolic background at 100 K is shown as color-contour maps in Figs. 2(a)–2(c). In all three crystals, with decreasing temperature, there is a suppression around zero bias below 60–70 K and an enhancement around $-20$ mV, indicative of the formation of a bulk HG [28] in agreement with an ARPES study [22, 27]. Thus, independent of the Sm-deficiency level, a HG of comparable size ($\Delta_{hyb}=21$ meV) opens at approximately the same temperature, explaining the similarly insulating behavior in $R_n(T)$ down to $\sim$10 K. Since the bulk HG is an essential prerequisite for the TSSs in TKIs [7], its robustness against Sm vacancies suggests that the TSSs may still be responsible for the slowdown in FZs. But, then, why doesn't $R_n(T)$ form a clear plateau as in the FG? To address this, we first scrutinize $g(V)$ curves at the lowest temperature [Figs. 2(d)–2(f)]. Here, two curves are shown for each crystal, one with the Pb superconducting and the other with it driven normal by applied magnetic fields. In all cases, they overlap well at high bias, as expected, including the broad peak around $-20$ mV. At low bias, they deviate drastically: sharp Pb coherence peaks (Pb superconducting) vs. an overall V-shape (Pb non-superconducting), the latter being reminiscent of the density of states (DOS) for Dirac fermions. Such V-shaped, *i.e.*, linear conductance has been reported in STS studies of $Bi_2Se_3$ [40, 41] but is missing in STS [15, 24, 25] and PCS [26] data on $SmB_6$. Possible reasons for these discrepant observations are discussed in Sec. 4 of SM [27]. Notice an additional peak at 5–6 mV appearing *only* in the positive bias branch. Together with the coherence peaks being *asymmetric*, this was a crucial clue to unraveling the topological nature of $SmB_6$ [28]. Here, we demonstrate that the fine spectroscopic differences among the three crystals indicate the sensitiveness of the TSSs to the Sm deficiency.

Normalized $g(V)$ curves with the Pb driven normal are compared in Fig. 3(a) in a low-bias region. Two common features appear in the overall V-shaped curves: A broad hump around $-2$ mV and a kink at $+4$ mV. Since this hump-kink structure originates from the TSSs interacting with the SEs [28], their common observation suggests such interaction still exists in the FZs. A clear difference is also observed: On the positive bias side, there are two distinct slopes for the FG whereas the curve is just quasi-linear for the FZs. The former originates from two distinguishable surface Dirac cones centered at $\bar{\Gamma}$ and $\bar{X}$ points in the (001) surface Brillouin zone [28]. Thus, the latter nominally indicates the two Dirac cones are indistinguishable in the FZs, presumably due to their intermixing. To find the microscopic origin, the conductance at each temperature with the Pb superconducting is normalized by that with it driven normal [Figs. 3(b)-3(d)]. Such normalization effectively divides out the spectral density in SmB$_6$ and, thus, would leave only the superconducting Pb features behind. However, additional features are revealed, e.g., in the FG [Fig. 3(b)]: (i) the asymmetry in the coherence peaks (marked by V+ and V-) increases with increasing temperature, and (ii) the peak labeled as V$_1$ is more clearly visible. These features were shown to originate from the inelastic tunneling involving SEs in SmB$_6$ [28]. The V$_1$ peak's Pb phononic origin can be completely ruled out by its independence of the counter-electrode (Sec. 5 in [27]). There are important contrasts among the crystals: (i) In the FZs, the V$_1$ peak is not as pronounced as in the FG despite the similar temperature dependence of its normalized height [Fig. 3(e)], and (ii) The coherence peaks in the FZs remain asymmetric down to the lowest temperature, whereas they become symmetric below ~4 K in the FG, as shown more quantitatively in Fig. 3(f) by the normalized peak height getting close to 1 below ~4 K (see the inset).

The inelastic tunneling features arising from processes occurring in the SmB$_6$ are more clearly visible when the Pb is superconducting because of the sharp coherence peaks in its single particle DOS, thereby increasing the energy resolution (note the tunneling spectrum is a convolution of the Pb and SmB$_6$ DOS) [28]. The V$_1$ peak originates from the emission of SEs at eV=$\Delta_{Pb}$+$\omega_0$, where $\omega_0$ is the characteristic SE energy and $\Delta_{Pb}$ is the superconducting gap energy, at which the *occupied* DOS (i.e., where electrons tunnel from) is peaked. Meanwhile, the contribution from the absorption channel is most pronounced at eV=$-\Delta_{Pb}$ owing to the peaked *empty* DOS (i.e., where electrons tunnel into), leading to the coherence-peak asymmetry. Thus, the persistence of the V$_1$ peak in FZs indicates the SEs still exist at the lowest measurement temperature. When the Pb is driven normal, the same inelastic tunneling effect must still exist albeit much weakened due to the lower energy resolution. However, such inelastic tunneling cannot account for the hump-kink structure observed when the Pb is driven normal since, due to the flat DOS, the emission process should cause only a parallel shift of the conductance for eV$\geq\omega_0$ and the absorption process can occur without any characteristic energy scale. This indicates the hump-kink structure appears via elastic tunneling. Thus, it is an intrinsic characteristic of the surface spectral density, modified via the electronic self-energy change due to the exchange of *virtual* SEs [34], analogously to the hump-dip structure in Pb DOS arising from the exchange of virtual phonons [42].

To reveal the underlying microscopic origin for the discrepant behaviors among the three crystals, the conductance is measured at a fixed bias voltage ($V_b$) over a wide temperature range from 1.8 K to 100 K, $g(V_b, T) \equiv \frac{dI}{dV}(T)\big|_{V=V_b}$. It is then normalized against $V_b = -50$ mV, namely, $g_n(V_b, T) \equiv g(V_b, T)/g(-50\text{mV}, T)$. Figure 4 displays $g_n(V_b, T)$ for two characteristic voltages: $-21$ mV and $+4$ mV. For $V_b = -21$ mV, they commonly exhibit a turning point around 60–70 K, indicative of the opening of a HG, in accordance with Figs. 2(a)–(c). However, clear differences appear at low temperature: A rapid increase followed by plateauing below ~4 K for the FG vs. gradual changes for FZs. For $V_b = +4$mV, in contrast to the FZs, the FG conductance decreases monotonically down to 15–20 K, below which it turns up. This temperature corresponds to the resistive hump (Fig. 1) and hence its detailed analysis may help uncover the origin of the two activation gaps. If $SmB_6$ were a topologically trivial KI, the conductance within the HG would rapidly decrease to zero with decreasing temperature. Therefore, the low-temperature upturn must signify the emergence of the TSSs. Upon further lowering temperature, $g_n(+4\text{mV}, T)$ shows distinct changes among the three crystals, similarly to $g_n(-21\text{mV}, T)$. Again, these results show spectroscopic properties of the TSSs are quite different among the crystals despite the similarity in the bulk gap, as detailed below.

## III. DISCUSSION

Our tunneling data reveal strong evidence for the interaction of the TSSs in $SmB_6$ with the SEs [28]. Then, the intriguing temperature dependence of both $g_n(+4\text{mV}, T)$ and $R_n(T)$ might be caused by that of the SEs. The impact would be drastic at low temperature if, being a bosonic excitation, they undergo a Bose-Einstein condensation. Within this scenario, the rapid increase in $g_n(+4\text{mV}, T)$ followed by plateauing might indicate the TSSs in the FG become coherent as the SEs freeze out or condense. This is also hinted by the Pb coherence peaks becoming symmetric below ~4 K [Figs. 3(b) and 3(f)]. This is because, as the SEs condense, the aforementioned inelastic tunneling channel involving the absorption of SEs must be suppressed. In strong contrast to the FG, the TSSs in FZs don't exhibit such a distinct change in $g_n(+4\text{mV}, T)$ because the SEs remain uncondensed. Namely, the TSSs continue to interact with them and, thus, stay incoherent down to the lowest measurement temperature. These observations provide an answer to our earlier question: The low-temperature resistive slowdown behavior in FZs (Fig. 1) is still due to the TSSs rather than the in-gap states; however, $R_n(T)$ does not exhibit a clear plateau because the TSSs remain incoherent due to their continued interaction with the SEs. The condensation of SEs invoked here has also been reported in a muon spin rotation study on flux-grown SmB$_6$ crystals [43]. In addition, according to Valentine *et al.*'s report from Raman scattering spectroscopy on three SmB$_6$ crystals grown in similar batches to ours [44], the SE peak is sharp in the stoichiometric FG but becomes broader or even missing in Sm-deficient FZs despite the similar HG size as observed in our tunneling data. Thus, while the HG itself is not sensitive to low-level Sm deficiency, the properties of the SEs change drastically. Consequently, crystals with the Sm deficiency differing by as small amount as 1% exhibit quite different behaviors in surface transport properties, in strong contrast to their weakly correlated counterparts [45]. Considering the growing interest in exciton physics on various condensed-matter platforms including graphene [46, 47] and transition-metal dichalcogenides [48], it is of great interest to further investigate the condensation of SEs in SmB$_6$. For instance, the measurement of $d^2I/dV^2$ may give the SE spectrum, analogously to the pairing-involved phonon spectrum in Pb [42]. The temperature dependence of such a spectrum may reveal the details underlying the exciton condensation process.

The sensitiveness of the topological nature to the Sm deficiency must affect many other properties in SmB$_6$. In particular, to resolve the enduring controversy over the origin of quantum oscillations in SmB$_6$ [9-12], more systematic studies should be carried out with this aspect considered carefully. For instance, it is worthwhile to see how the signal indicative of the surface states [9] would change with the Sm deficiency. For a more microscopic understanding of our results, further theoretical investigation is necessary to study how the surface spectral density is modified due to the interaction with the SEs [34] in the presence of disorder, in particular, the Sm deficiency [49, 50]. The persistence of the HG up to 1% of Sm deficiency indicates that the translational invariance in the Kondo lattice, essential to the formation of coherent (hybridized) bands, may not yet be broken globally, still allowing the TSSs to emerge, albeit with quite different properties.

## IV. CONCLUSION

In conclusion, our comparative tunneling studies on SmB$_6$ crystals show that the TSSs still exist for up to 1% of Sm deficiency, in line with the persistent bulk HG. However, their temperature evolution is distinctively different depending on the deficiency level as their interaction with the SEs varies drastically. In a stoichiometric crystal, the TSSs become coherent as the SEs condense, causing the resistance to exhibit a clear plateau at low temperature. In sharp contrast, the TSSs remain incoherent in Sm-deficient crystals due to their continued interaction with the SEs, thus showing a non-saturating resistive behavior. This disorder-sensitiveness of the topological nature in $SmB_6$, very likely TKIs in general, should be carefully considered when addressing the questions that remain open after a decade of intensive research. Theoretically, it will be important for future work to investigate how the influence of the SEs on the surface spectral density is modified by the Sm deficiency.

## ACKNOWLEDGEMENTS

The wok at FSU and NHMFL was supported under Award No. NSF/DMR-1704712, and in part under Award No. NSF/DMR-1644779 and the State of Florida. The work at JHU was supported under Award No. DOE/BES EFRC DE-SC0019331. TMM acknowledges JHU Catalyst Fund.

**FIGURE CAPTIONS**

**Figure 1. Normalized DC electrical resistance measured across the (001) surface of three SmB$_6$ crystals**. Double-logarithmic (main panel) and Arrhenius plot (inset) of the normalized resistance, R$_n$(T) ≡ R(T)/R(300K). While they exhibit similarly insulating behaviors at high temperature, a large discrepancy is apparent below ~10 K.

**Figure 2. Bulk HG and the TSSs in SmB$_6$.** (a-c) Color-contour maps of the normalized tunneling conductance, $g$(V). Below the $T_c$ (7.2K) of Pb, conductance was measured with the Pb driven normal by magnetic fields indicated in (d-f). Overall, all three crystals exhibit similar features due to the opening of a bulk HG. (d-f) Normalized conductance at the lowest temperature with the Pb superconducting (blue solid lines) and driven normal (red solid lines). An additional peak is seen only in the positive-bias branch (5 – 6 mV).

**Figure 3. Influence of the SEs on the TSSs.** (a) Elastic tunneling effect when the Pb is driven normal. Normalized $g$(V) shows common features of hump (around −2 mV) and kink (~4 mV) originating from the interaction with virtual SEs. An apparent discrepancy is two linear slopes (FG) vs. quasilinear shape (FZs), possibly due to intermixing of the two surface Dirac cones. (b-d) Inelastic tunneling effect is clearly seen when $g$(V) with the Pb superconducting is divided out by $g$(V) when the Pb driven normal by applied magnetic fields. (e) V$_1$ peak height normalized against the positive-bias coherence peak (V+). (f) Height of the negative-bias coherence peak (V-) normalized against V+. Inset: a zoomed view for the low-temperature region.

**Figure 4. Temperature evolution of the TSSs.** (a-c) Temperature dependence of conductance at two fixed bias voltages ($V_b$ = −21 mV and +4 mV) normalized against −50 mV. At low temperature, while the FG exhibits a rapid increase followed by plateauing, FZs exhibit gradual changes.

## REFERENCES

[1]     X. L. Qi and S. C. Zhang, Rev. Mod. Phys. **83**, 1057 (2011).

[2]     M. Z. Hasan and J. E. Moore, Annu. Rev. Condens. Matter Phys. **2**, 55 (2011).

[3]     L. Fu, C. L. Kane and E. J. Mele, Phys. Rev. Lett. **98**, 106803 (2007).

[4]     P. S. Riseborough, Adv. Phys. **49**, 257 (2000).

[5]     P. Coleman, Handbook of Magnetism and Advanced Magnetic Materials (Wiley, New York, 2007), pp. 95-148.

[6]     M. Dzero, J. Xia, V. Galitski and P. Coleman, Annu. Rev. Condens. Matter Phys. **7**, 249 (2016).

[7]     M. Dzero, K. Sun, V. Galitski and P. Coleman, Phys. Rev. Lett. **104**, 106408 (2010).

[8]     J. W. Allen, Philos. Mag. **96**, 3227 (2016).

[9]     G. Li, Z. Xiang, F. Yu, T. Asaba, B. Lawson, P. Cai, C. Tinsman, A. Berkley, S. Wolgast, Y. S. Eo, D. J. Kim, C. Kurdak, J. W. Allen, K. Sun, X. H. Chen, Y. Y. Wang, Z. Fisk and L. Li, Science **346**, 1208 (2014).

[10]    Z. Xiang, B. Lawson, T. Asaba, C. Tinsman, L. Chen, C. Shang, X. H. Chen and L. Li, Phys. Rev. X **7**, 031054 (2017).

[11]    B. S. Tan, Y. T. Hsu, B. Zeng, M. C. Hatnean, N. Harrison, Z. Zhu, M. Hartstein, M. Kiourlappou, A. Srivastava, M. D. Johannes, T. P. Murphy, J. H. Park, L. Balicas, G. G. Lonzarich, G. Balakrishnan and S. E. Sebastian, Science **349**, 287 (2015).

[12]    S. M. Thomas, X. X. Ding, F. Ronning, V. Zapf, J. D. Thompson, Z. Fisk, J. Xia and P. F. S. Rosa, Phys. Rev. Lett. **122**, 166401 (2019).

[13]    W. A. Phelan, S. M. Koohpayeh, P. Cottingham, J. A. Tutmaher, J. C. Leiner, M. D. Lumsden, C. M. Lavelle, X. P. Wang, C. Hoffmann, M. A. Siegler, N. Haldolaarachchige, D. P. Young and T. M. McQueen, Sci. Rep. **6**, 20860 (2016).

[14]    W. T. Fuhrman, J. R. Chamorro, P. A. Alekseev, J. M. Mignot, T. Keller, J. A. Rodriguez-Rivera, Y. Qiu, P. Nikolic, T. M. McQueen and C. L. Broholm, Nat. Commun. **9**, 1539 (2018).

[15]    M. M. Yee, Y. He, A. Soumyanarayanan, D.-J. Kim, Z. Fisk, and J. E. Hoffman, arXiv:1308.1085.

[16]    Z. H. Zhu, A. Nicolaou, G. Levy, N. P. Butch, P. Syers, X. F. Wang, J. Paglione, G. A. Sawatzky, I. S. Elfimov and A. Damascelli, Phys. Rev. Lett. **111**, 216402 (2013).

[17]    J. Denlinger, J. W. Allen, J.-S. Kang, K. Sun, J.-W. Kim, J. H. Shim, B. I. Min, D.-J. Kim and Z. Fisk, arXiv:1312.6637.

[18]    E. Frantzeskakis, N. de Jong, B. Zwartsenberg, Y. K. Huang, Y. Pan, X. Zhang, J. X. Zhang, F. X. Zhang, L. H. Bao, O. Tegus, A. Varykhalov, A. de Visser and M. S. Golden, Phys. Rev. X **3**, 041024 (2013).

[19]    J. Jiang, S. Li, T. Zhang, Z. Sun, F. Chen, Z. R. Ye, M. Xu, Q. Q. Ge, S. Y. Tan, X. H. Niu, M. Xia, B. P. Xie, Y. F. Li, X. H. Chen, H. H. Wen and D. L. Feng, Nat. Commun. **4**, 3010 (2013).

[20]    C. H. Min, P. Lutz, S. Fiedler, B. Y. Kang, B. K. Cho, H. D. Kim, H. Bentmann and F. Reinert, Phys. Rev. Lett. **112**, 226402 (2014).

[21]    H. Miyazaki, T. Hajiri, T. Ito, S. Kunii and S. I. Kimura, Phys. Rev. B **86**, 075105 (2012).

[22]    M. Neupane, N. Alidoust, S. Y. Xu, T. Kondo, Y. Ishida, D. J. Kim, C. Liu, I. Belopolski, Y. J. Jo, T. R. Chang, H. T. Jeng, T. Durakiewicz, L. Balicas, H. Lin, A. Bansil, S. Shin, Z. Fisk and M. Z. Hasan, Nat. Commun. **4**, 2991 (2013).

[23]    N. Xu, P. K. Biswas, J. H. Dil, R. S. Dhaka, G. Landolt, S. Muff, C. E. Matt, X. Shi, N. C. Plumb, M. Radovic, E. Pomjakushina, K. Conder, A. Amato, S. V. Borisenko, R. Yu, H. M. Weng, Z. Fang, X. Dai, J. Mesot, H. Ding and M. Shi, Nat. Commun. **5**, 4566 (2014).

[24]    S. Rössler, T. H. Jang, D. J. Kim, L. H. Tjeng, Z. Fisk, F. Steglich and S. Wirth, Proc. Natl. Acad. Sci. U.S.A. **111**, 4798 (2014).

[25]    W. Ruan, C. Ye, M. H. Guo, F. Chen, X. H. Chen, G. M. Zhang and Y. Y. Wang, Phys. Rev. Lett. **112**, 136401 (2014).

[26]    X. H. Zhang, N. P. Butch, P. Syers, S. Ziemak, R. L. Greene and J. Paglione, Phys. Rev. X **3**, 011011 (2013).

[27]    See Supplemental Material at http://link.aps.org/supplemental/10.1103/PhysRevB.103. 155125 for the discussion of materials and methods (Sec. 1); characterization of the tunnel barrier (Sec. 2); comparison with photoemission results (Sec. 3); comparison with scanning tunneling and point-contact spectroscopies (Sec. 4); and independence of inelastic tunneling features on the counter-electrode (Sec. 5).

[28]  W. K. Park, L. Sun, A. Noddings, D.-J. Kim, Z. Fisk and L. H. Greene, Proc. Natl. Acad. Sci. U.S.A. **113**, 6599 (2016).

[29]  F. Lu, J. Z. Zhao, H. M. Weng, Z. Fang and X. Dai, Phys. Rev. Lett. **110**, 096401 (2013).

[30]  V. Alexandrov, M. Dzero and P. Coleman, Phys. Rev. Lett. **111**, 226403 (2013).

[31]  J. Kim, K. Kim, C. J. Kang, S. Kim, H. C. Choi, J. S. Kang, J. D. Denlinger and B. I. Min, Phys. Rev. B **90**, 075131 (2014).

[32]  T. Takimoto, J. Phys. Soc. Jpn. **80**, 123710 (2011).

[33]  W. T. Fuhrman, J. Leiner, P. Nikolic, G. E. Granroth, M. B. Stone, M. D. Lumsden, L. DeBeer-Schmitt, P. A. Alekseev, J. M. Mignot, S. M. Koohpayeh, P. Cottingham, W. A. Phelan, L. Schoop, T. M. McQueen and C. Broholm, Phys. Rev. Lett. **114**, 036401 (2015).

[34]  G. A. Kapilevich, P. S. Riseborough, A. X. Gray, M. Gulacsi, T. Durakiewicz and J. L. Smith, Phys. Rev. B **92**, 085133 (2015).

[35]  A. Arab, A. X. Gray, S. Nemsak, D. V. Evtushinsky, C. M. Schneider, D. J. Kim, Z. Fisk, P. F. S. Rosa, T. Durakiewicz and P. S. Riseborough, Phys. Rev. B **94**, 235125 (2016).

[36]  B. Gorshunov, N. Sluchanko, A. Volkov, M. Dressel, G. Knebel, A. Loidl and S. Kunii, Phys. Rev. B **59**, 1808 (1999).

[37]  P. Schlottmann, Phys. Rev. B **46**, 998 (1992).

[38]  L. Sun, D. J. Kim, Z. Fisk and W. K. Park, Phys. Rev. B **95**, 195129 (2017).

[39]  D. J. Kim, J. Xia and Z. Fisk, Nat. Mater. **13**, 466 (2014).

[40]  P. Cheng, C. L. Song, T. Zhang, Y. Y. Zhang, Y. L. Wang, J. F. Jia, J. Wang, Y. Y. Wang, B. F. Zhu, X. Chen, X. C. Ma, K. He, L. L. Wang, X. Dai, Z. Fang, X. C. Xie, X. L. Qi, C. X. Liu, S. C. Zhang and Q. K. Xue, Phys. Rev. Lett. **105**, 076801 (2010).

[41]  T. Hanaguri, K. Igarashi, M. Kawamura, H. Takagi and T. Sasagawa, Phys. Rev. B **82**, 081305(R) (2010).

[42]  W. L. McMillan and J. M. Rowell, Phys. Rev. Lett. **14**, 108 (1965).

[43]  K. Akintola, A. Pal, S. R. Dunsiger, A. C. Y. Fang, M. Potma, S. R. Saha, X. F. Wang, J. Paglione and J. E. Sonier, npj Quantum Mater. **3**, 36 (2018).

[44]  M. E. Valentine, S. Koohpayeh, W. A. Phelan, T. M. McQueen, P. F. S. Rosa, Z. Fisk and N. Drichko, Physica B **536**, 60 (2018).

[45]     J. G. Checkelsky, Y. S. Hor, M. H. Liu, D. X. Qu, R. J. Cava and N. P. Ong, Phys. Rev. Lett. **103**, 246601 (2009).

[46]     J. I. A. Li, T. Taniguchi, K. Watanabe, J. Hone and C. R. Dean, Nat. Phys. **13**, 751 (2017).

[47]     X. M. Liu, K. Watanabe, T. Taniguchi, B. I. Halperin and P. Kim, Nat. Phys. **13**, 746 (2017).

[48]     Z. F. Wang, D. A. Rhodes, K. Watanabe, T. Taniguchi, J. C. Hone, J. Shan and K. F. Mak, Nature **574**, 76 (2019).

[49]     P. S. Riseborough, Phys. Rev. B **68**, 235213 (2003).

[50]     M. Abele, X. Yuan and P. S. Riseborough, Phys. Rev. B **101**, 094101 (2020).

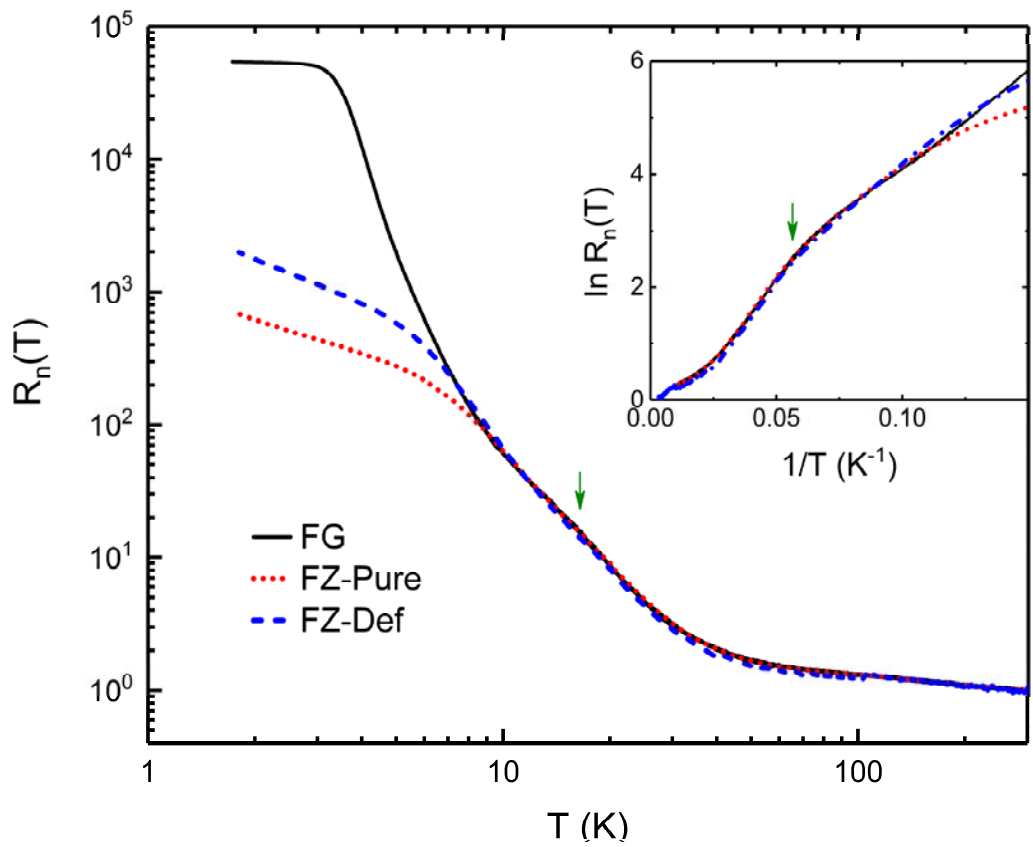

**Figure 1**

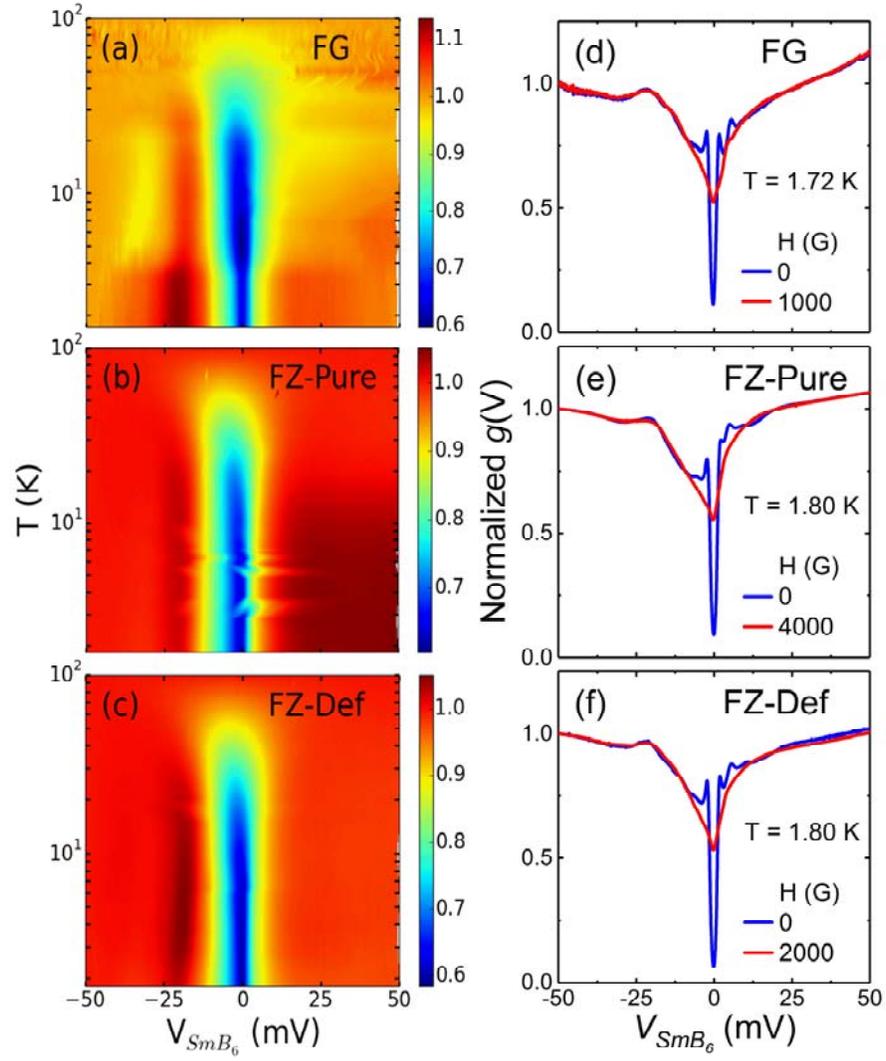

**Figure 2**

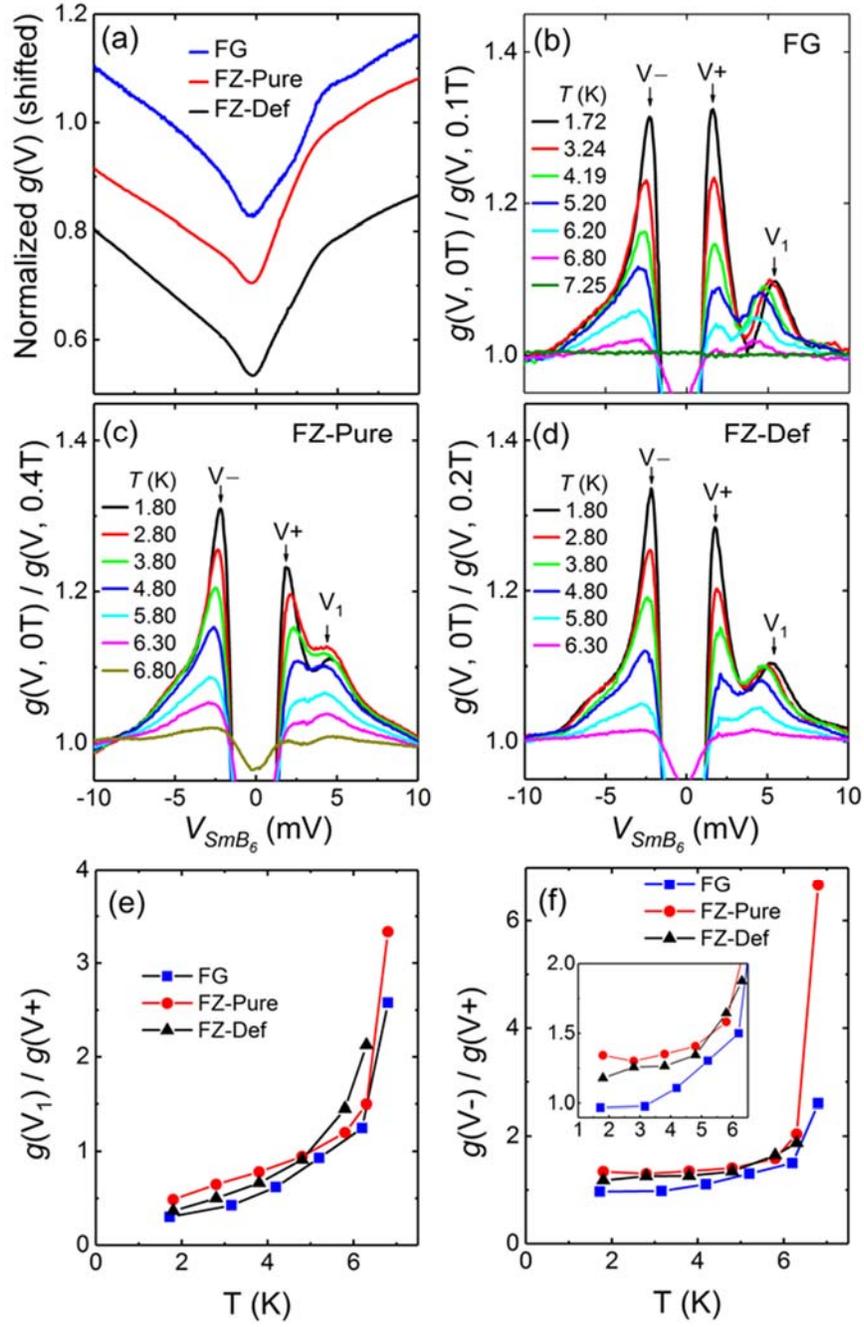

**Figure 3**

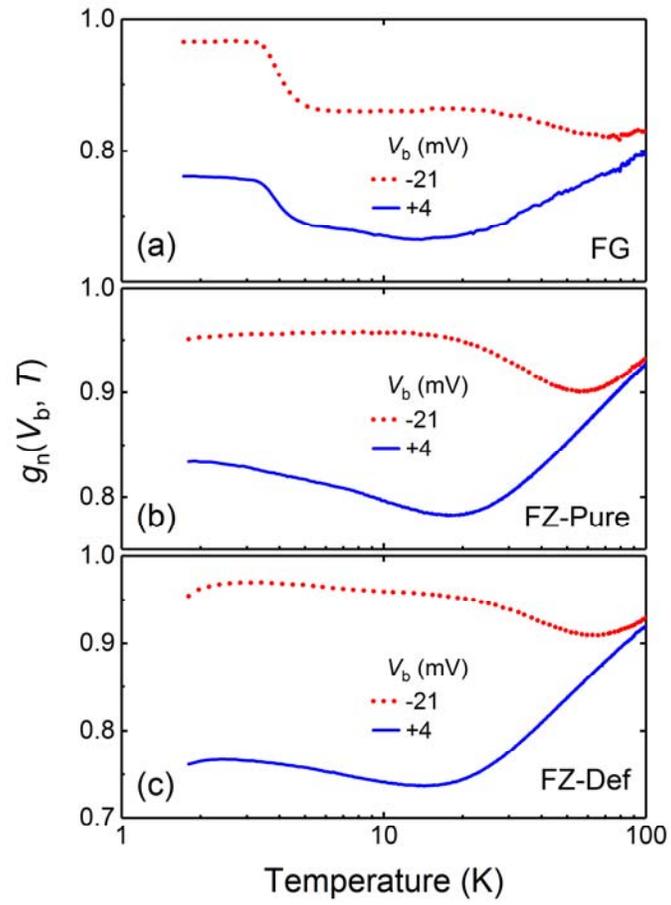

**Figure 4**

**Supplemental Material**

**Topological nature of the Kondo insulator $SmB_6$ and its sensitiveness to Sm vacancy**

W. K. Park[1,*], J. A. Sittler,[1,2,a] L. H. Greene,[1,2] W. T. Fuhrman,[3] J. Chamorro,[3] S. Koohpayeh,[3] W. Adam Phelan,[3] T. M. McQueen[3]

*[1]National High Magnetic Field Laboratory, Florida State University, Tallahassee, FL 32310, USA*
*[2]Department of Physics, Florida State University, Tallahassee, FL 32306, USA*
*[3]Department of Physics and Astronomy, Hopkins University, Baltimore, MD 21210, USA*

**Table of Contents**

[*]Corresponding author, wkpark@magnet.fsu.edu
[a]Present address: Department of Physics, University of California Santa Cruz

# 1. Materials and methods

SmB$_6$ single crystals were grown by Al-flux [1] and floating-zone methods [2]. Their crystallographic axes are identified by single-crystal x-ray diffraction. Based on this information, the crystals are cut such that the surface orientation is along the (001) and (011) axes, respectively. For tunnel junctions, the cut crystals are embedded in epoxy molds (Stycast® 2850-FT) for easiness in handling and polishing. The crystals are polished using alumina lapping films with particle size in the range of $12 - 0.3$ μm. The polished surfaces are extremely smooth with an average peak-to-dip distance of $0.4 - 0.8$ nm as measured with an atomic force microscope. Polished crystals are loaded into a high-vacuum chamber for the formation of a tunnel barrier. High-quality tunnel junctions are formed by oxidizing the polished crystal surface that is cleaned *in situ* with Ar ion beam. The barrier layer is confirmed to be B$_2$O$_3$ by x-ray photoemission and Auger electron spectroscopies [3]. The crystal edges are painted with diluted Duco® cement for insulation purpose prior to the deposition of the counter-electrode (Pb). Thin Pb strips are evaporated through a shadow mask.

The finished sample is mounted on a sample puck for a Quantum Design PPMS and aluminum strips are attached for contact leads. Electrical resistance is measured with a standard four-probe technique with a DC current source and a nanovoltmeter. Differential conductance across a tunnel junction is measured using a standard lock-in technique. Schematic cross-sectional views for both resistance and tunneling conductance measurements are shown in Fig. S1a. Figure S1b compares resistance vs. temperature curves taken before and after the plasma oxidation. Overall, the two curves are almost identical including the low-temperature plateau. This indicates that the surface states that existed at the top surface now have moved down to underneath the newly formed surface B$_2$O$_3$ layer. It is these moved surface states that are detected in our tunneling conductance, as discussed in the main text. In summary, the combination of adopted processes in our PTS, particularly the formation of a robust tunnel barrier via self-oxidation, enables to reveal intrinsic spectroscopic properties of SmB$_6$.

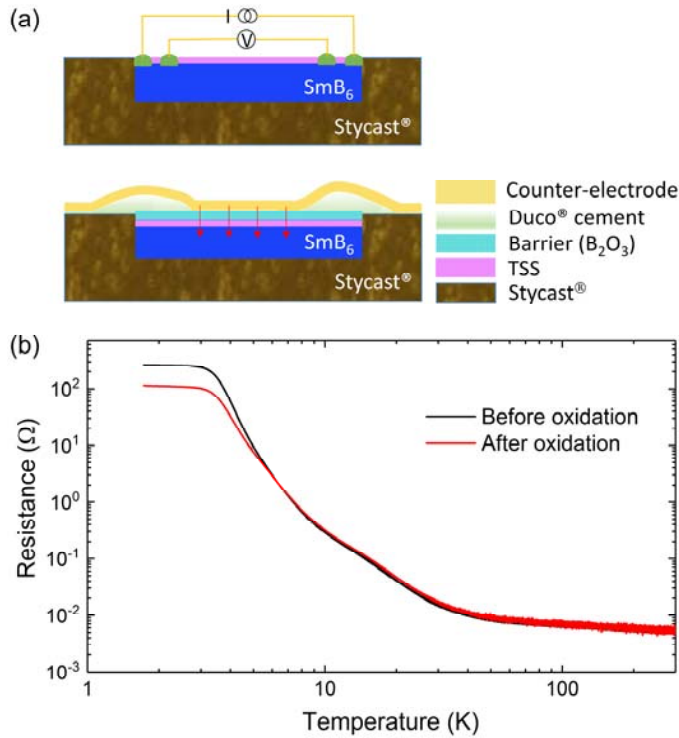

**Figure S1. Cross-sectional measurement configuration and comparison of resistance vs. temperature before and after oxidation.** (a) Cross-sectional schematic of the measurement configuration. Top panel: Four-probe resistance measurement. Bottom panel: Tunneling conductance across a SmB$_6$/SmB$_6$-Oxi/Pb junction. TSS denotes the topological surface states. In contrast to the resistance measurement, the confined geometry along with the surface-sensitiveness of PTS enables to associate the characteristic features at low bias with the surface states. (b) Comparison of resistance vs. temperature curves before and after the plasma oxidation of a flux-grown (001) SmB$_6$ crystal. The contact leads were made manually, so their dimensions/configuration is slightly different in the two measurements. All characteristics including the low-temperature plateau are virtually identical.

## 2. Characteristics of the tunnel barrier in SmB₆ junctions

Key properties of a suitable tunnel barrier include high-enough electrostatic potential and sharp interfaces with both bottom and top electrodes. A standard model to be used for the characterization of a tunnel barrier is by Brinkman, Dynes, and Rowell (BDR) [4]. Figure S2 displays the differential conductance of a high-quality junction along with a best-fit curve to the BDR model. The fit parameters for the barrier are: the thickness $\cong 10.0$ Å, the height $\cong 11.9$ eV, and the asymmetry $= 0.8$ eV. Apparently, the deduced potential is quite high and falls in the range of a calculated band gap [5].

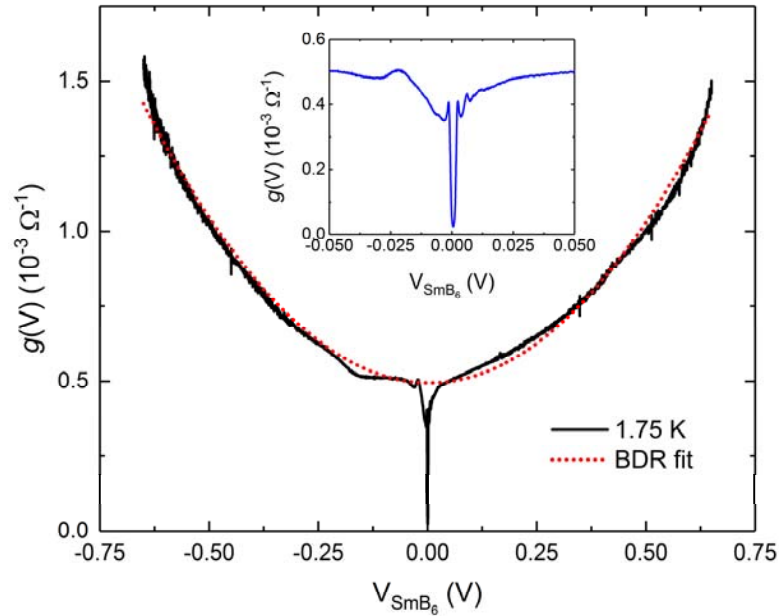

**Figure S2.** Tunneling conductance taken at 1.75 K of a high-quality SmB₆/SmB₆-Oxi/Pb junction (black solid line) and the best-fit curve of the parabolic background to the BDR model (red dotted line) [4]. The extracted barrier parameters are: the thickness $\cong 10.0$ Å, the potential barrier height $\cong 11.9$ eV, and the asymmetry $= 0.8$ eV, indicative of a high-quality tunnel barrier. This is also evidenced in the inset, where the same conductance is zoomed in the low-bias region. All features characteristic of a high-quality tunnel junction are observed including the broad peak around -21 mV, the asymmetric Pb coherence peaks, and the extra peak around 5 mV arising from inelastic tunnelling involving spin excitons (see main text).

## 3. Comparison with photoemission results

While the reports from several ARPES studies on SmB₆ lack a general agreement among themselves [6-12], one ARPES study [11] reveals features similar to those seen in our tunneling data. As shown in Fig. S3, both measurements show a peak due to the bulk gap at a similar energy scale, i.e., around −19 meV at 6 K. The authors interpreted the hump appearing close to (below) the Fermi level as a signature for the in-gap states. Considering the similarity in its shape, energy scale, and temperature evolution to those in our conductance spectra, we believe that its origin is the same, namely, interaction of the surface states with the spin excitons. If the ARPES measurement had been carried out at a temperature as low as in our tunneling spectroscopy, the intensity would have shown a linear region close to the Fermi level, similarly to the tunneling conductance. In the positive energy range, the ARPES data don't reveal much information since there the energy levels are unoccupied, whereas our tunneling data exhibit fine details, namely, the linear conductance up to ~4 mV followed by a kink. This is possible because the tunneling can occur in either direction depending on the bias polarity. While our tunneling conductance exhibits the hump-kink structure commonly on both (001) and (011) surfaces, it also reveals differences: two Dirac cones for (001) surface and one Dirac cone for (011) surface, as reported previously [13] and described in the main text.

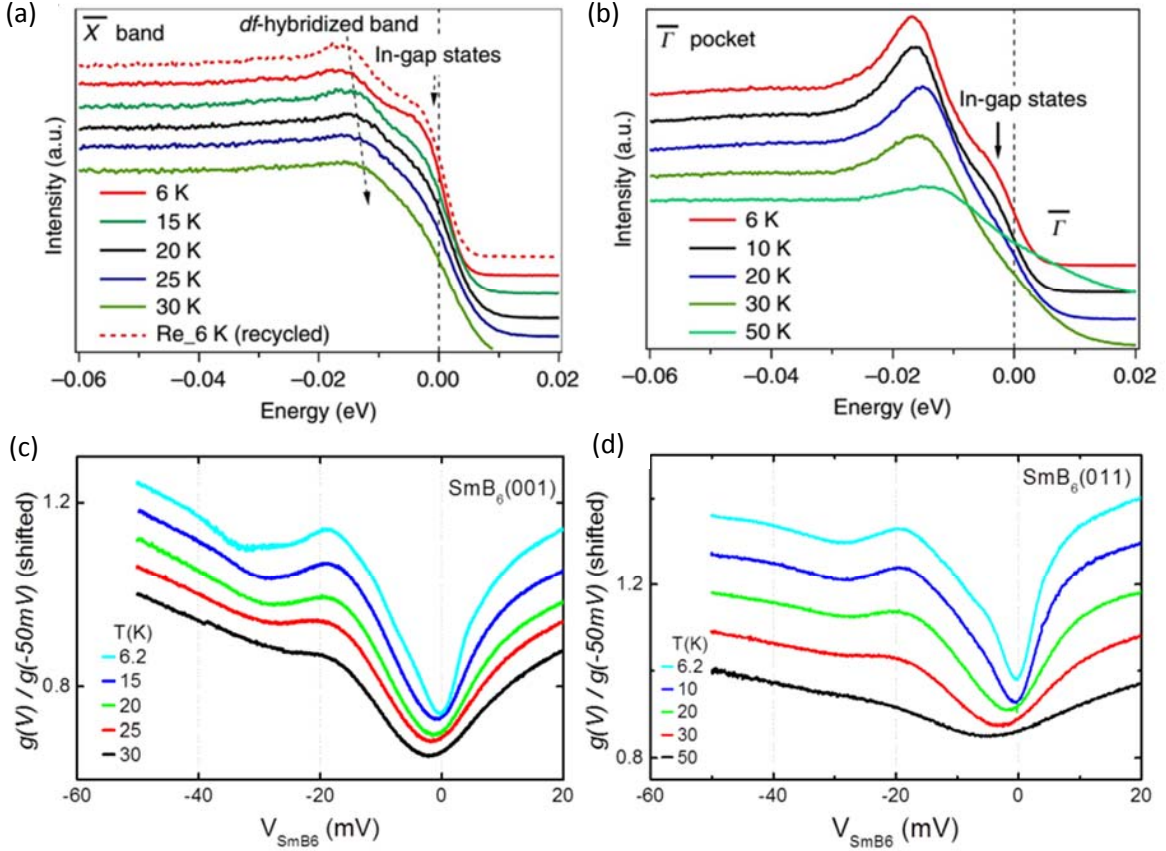

**Figure S3.** Comparison of one ARPES study [11] with our PTS. (a) & (b) ARPES data on a (001) SmB$_6$ surface, adapted from Ref. 11 with permission from the publisher. (c) & (d) Our tunneling conductance spectra plotted over the same energy scales and temperature ranges as in the ARPES data for comparison purpose. Notice the similarity in the bulk gap feature (broad peaks around −19 mV at 6 K) and the hump structure slightly below the Fermi level, which we interpret as due to the interaction with the spin excitons.

## 4. Comparison with scanning tunneling and point-contact spectroscopies

In sharp contrast with the ARPES result discussed in the previous section, both scanning tunneling spectroscopy (STS) and point-contact spectroscopy (PCS) provided conductance spectra with somewhat discrepant features from our planar tunneling data, as compared in Fig. S4. The tunneling conductance shown in Fig. S4d as reported in an STS study [14]) largely exhibits single-impurity-like Fano line shape [15] without showing clear evidence for the surface states. More specifically, STS spectra from SmB$_6$ are lacking the linear conductance at low bias, an expected signature for Dirac fermions as confirmed in STS studies on the topological insulator Bi$_2$Se$_3$ [16, 17], let alone the hump-kink structure seen in our tunneling data (Fig. S4b). In early STS studies [14, 18, 19], only the zero-bias conductance being finite was attributed to the surface states. The PCS data shown in Fig. S4e [20] look a bit different from the STS data in the sense that the overall shape is more reminiscent of the Fano resonance in a Kondo lattice rather than in a single Kondo impurity system. However, again clear signatures for the surface states are missing in the PCS data and the hybridization gap appears too large (~40 meV) unlike in the authors' report (18 meV) [20] and in the literature [21].

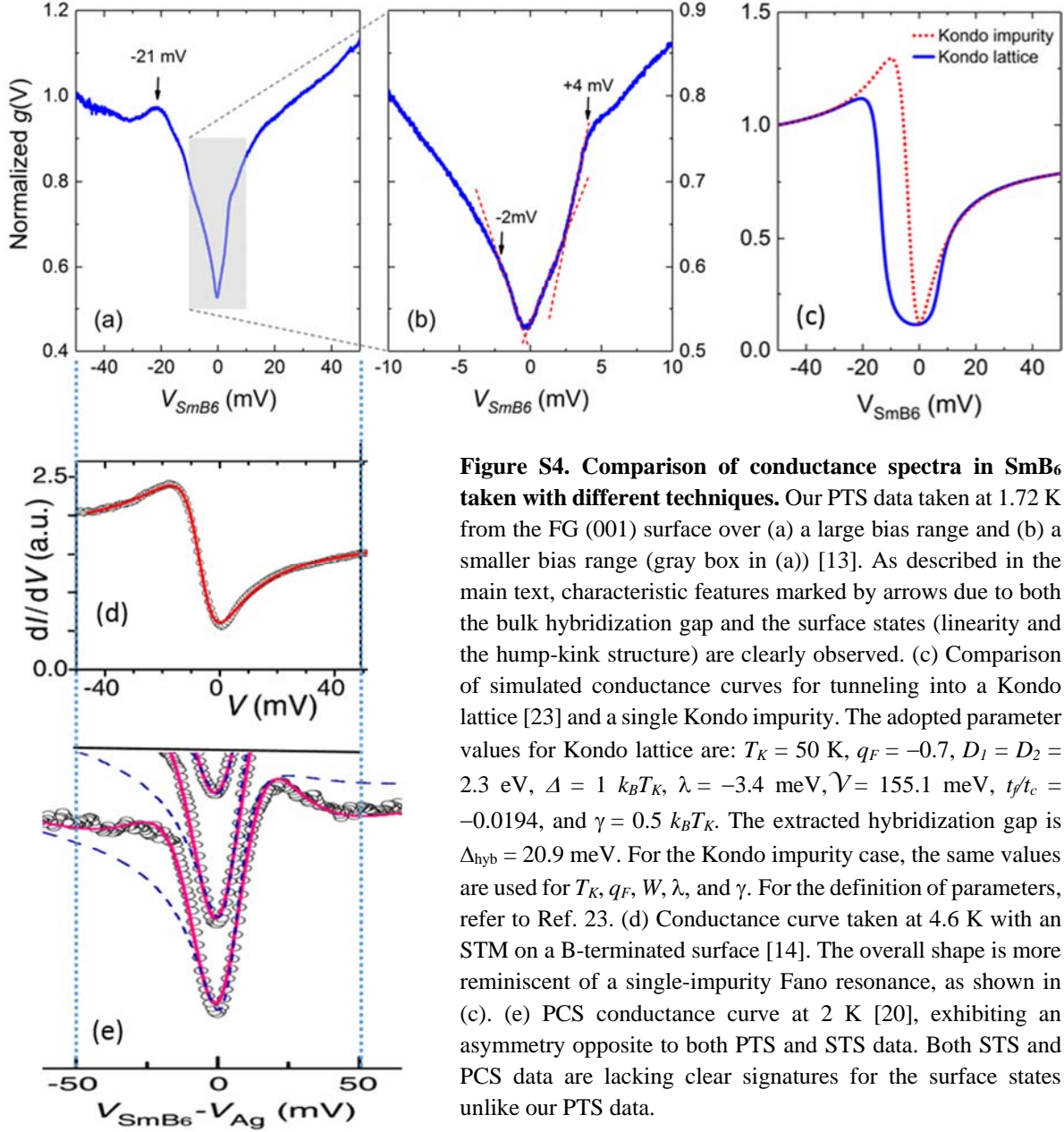

**Figure S4. Comparison of conductance spectra in SmB₆ taken with different techniques.** Our PTS data taken at 1.72 K from the FG (001) surface over (a) a large bias range and (b) a smaller bias range (gray box in (a)) [13]. As described in the main text, characteristic features marked by arrows due to both the bulk hybridization gap and the surface states (linearity and the hump-kink structure) are clearly observed. (c) Comparison of simulated conductance curves for tunneling into a Kondo lattice [23] and a single Kondo impurity. The adopted parameter values for Kondo lattice are: $T_K$ = 50 K, $q_F$ = −0.7, $D_1$ = $D_2$ = 2.3 eV, $\Delta$ = 1 $k_BT_K$, $\lambda$ = −3.4 meV, $V$ = 155.1 meV, $t_f/t_c$ = −0.0194, and $\gamma$ = 0.5 $k_BT_K$. The extracted hybridization gap is $\Delta_{hyb}$ = 20.9 meV. For the Kondo impurity case, the same values are used for $T_K$, $q_F$, $W$, $\lambda$, and $\gamma$. For the definition of parameters, refer to Ref. 23. (d) Conductance curve taken at 4.6 K with an STM on a B-terminated surface [14]. The overall shape is more reminiscent of a single-impurity Fano resonance, as shown in (c). (e) PCS conductance curve at 2 K [20], exhibiting an asymmetry opposite to both PTS and STS data. Both STS and PCS data are lacking clear signatures for the surface states unlike our PTS data.

It is generally agreed that the conductance measured with all three techniques share a common Fano resonance phenomenon [22] as manifested by the asymmetric line shape. However, it is crucial to distinguish detailed features observed in each configuration in order to understand the reason why these probes that are seemingly sensitive to the electronic density of states give such discrepant results. In Fig. S5, calculated conductance spectra are compared for tunneling into a single-Kondo-impurity system and a Kondo lattice using the Maltseva-Dzero-Coleman (MDC) model [23]. Despite the commonly seen asymmetry characteristic of a Fano resonance [22], a clear difference is also seen: While the conductance curve for a Kondo lattice shows a clear gap due to the hybridization, no gap is seen in the single impurity case as it shouldn't be. This contrasting feature is also shown in Fig. S4c, where simulation of the SmB₆ data is displayed for the two cases. While the Kondo impurity curve resembles the STS data (Fig. S4d), the Kondo lattice curve contains similar features to our tunneling data (Fig. S4a) including the broad peak at −20 mV and the extracted hybridization gap size of 20.9 meV. The deviation from the actual data is partly because the contribution from the surface states is not yet incorporated in this model. It'll be a worthwhile task to analyze our data more quantitatively with such a model in the future.

In order to understand the origin of the discrepancy between the STS and our PTS results, a crucial clue may lie in what's underlying the physics of topological Kondo insulators, in particular, the fact that apparently $SmB_6$ is a Kondo lattice system. Here, we note that theoretically the topological surface states emerge only if the bulk band structure exhibits a hybridization gap. In other words, the lattice coherence is essential, which is in turn rooted on the translational invariance that exists in a lattice but not in a single impurity system. Likewise, since the topological surface states are rooted on the inversion of the bulk bands due to large spin-orbit coupling, it is fairly likely that their emergence and characteristics also rely on the translational invariance. That being granted, it should be questioned whether an STM, a local probe on the subatomic scale, can pick up the bulk hybridization gap and the TSS that are both rooted on the global nature of the Kondo lattice. As the tip in an STM has to sit above an atom, the resulting conductance would naturally exhibit single-impurity-like Fano line shape, as seen in actual STS data [14, 18, 19]. Whether such data can still contain information on the hybridization gap and TSS needs to be investigated rigorously. This is a quite different situation from weakly correlated topological insulators, e.g., $Bi_2Se_3$, where several STS studies were able to quickly reveal the characteristics of the TSS such as the Dirac fermions' linear density of states and its quantization under magnetic fields [16, 17]. The non-trivial Berry phase extracted from these data comprises one crucial piece of evidence for their topological nature. In contrast, despite longstanding research on $SmB_6$ over the last several years, STS studies carried out by several groups worldwide have not yet brought the key information on its topological nature.

At this point, it is desirable to understood how our PTS junctions are able to reveal the detailed features for the TSS that are missing in STS and PCS data, including the linear spectral density and the hump-kink structure due to the interaction with the spin excitons. It is well known that the surface atoms in SmB$_6$ easily undergo reconstructions [19], which has been a major challenge in obtaining clean STS data. On the other hand, in our PTS, the polished surface is *in situ* cleaned with low-energy Ar ions and immediately oxidized in oxygen plasma to form a tunnel barrier (B$_2$O$_3$). As seen in Fig. S1b, the resistance data from the same crystal before and after the oxidation are almost identical including the low-temperature plateau. Combined with the fact that the junction made on such an oxidized surface shows sharp tunneling characteristics, one can easily infer that the surface states must have moved down to underneath the surface oxide (B$_2$O$_3$) layer (see the bottom panel in Fig. S1a). This means that the characteristic signatures in our tunneling conductance originate from the TSS that are re-formed beneath B$_2$O$_3$, hence, are not affected by any disorder such as reconstruction that might exist on a pristine polished surface. Therefore, if STM being a local probe as mentioned above were not an inherent drawback, this fundamental difference in the process, namely, whether the TSS are protected by the insulating B$_2$O$_3$ layer against any disorder (our PTS) or not (STS) may be the main reason why the STS data are largely lacking detailed information on the TSS, as discussed above.

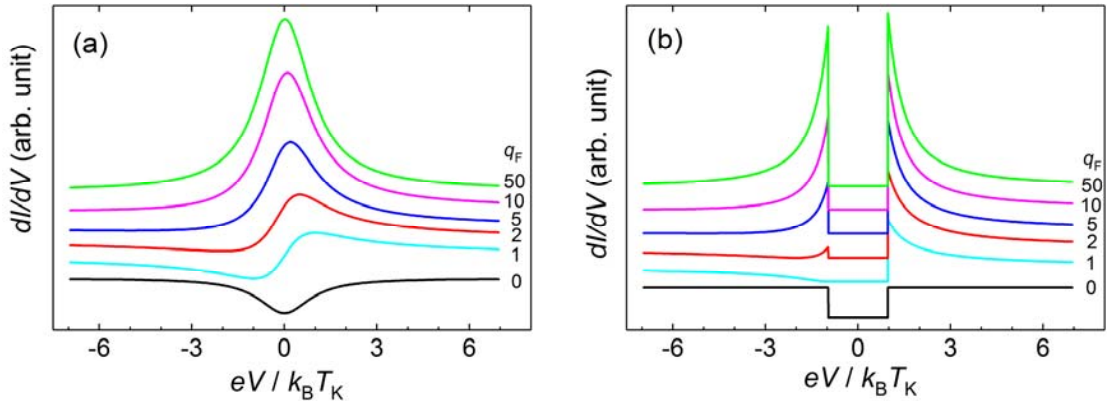

**Figure S5. Comparison of calculated conductance for tunneling into (a) a single-impurity Kondo system and (b) a Kondo lattice.** For the latter, the adopted model is by Maltseva-Dzero-Coleman. $T_K$ is a characteristic temperature for single impurity Kondo effect or lattice coherence. The parameter that governs the conductance shape in both cases is the Fano parameter, $q_F$, which is given by the ratio between two matrix amplitudes for tunneling into localized and itinerant states. While the conductance is symmetric for very small and large $q_F$ values, it is asymmetric for intermediate $q_F$ values due to an interference between the two channels. A fundamental difference between the two cases is that while a hybridization gap appears as two separate peaks in the Kondo lattice case, it doesn't exist in the single Kondo impurity system as it shouldn't.

## 5. Evidence for the independence of inelastic tunneling features on the counter-electrode

Although it was carefully ruled out in our previous study (see [13] including Supporting Information), one may still wonder whether the inelastic tunneling signatures taken as evidence for the involvement of spin excitons in SmB$_6$ are instead related to the phonon features in superconducting Pb used as the counter-electrode simply because the V$_1$ peak location is quite close to the hump-dip structure in Pb tunnel junctions. Here we provide a direct experimental evidence that such possibility can be excluded completely. Figure S6 compares conductance spectra taken from two junctions with the only difference being the counter-electrode: Pb and Sn. As it is known that the electron-phonon coupling in Sn is weak unlike in Pb, no phonon features are seen in the Al-AlO$_x$-Sn junction, in sharp contrast to the Al-AlO$_x$-Pb junction. However, as shown in Fig. S7, the V$_1$ peak is still observed albeit weak due to a larger thermal broadening effect in the Sn due to its lower T$_c$ (3.8 K) than the Pb (7.2 K). Therefore, one can rule out the Pb phonons as the source for inelastic tunneling in SmB$_6$.

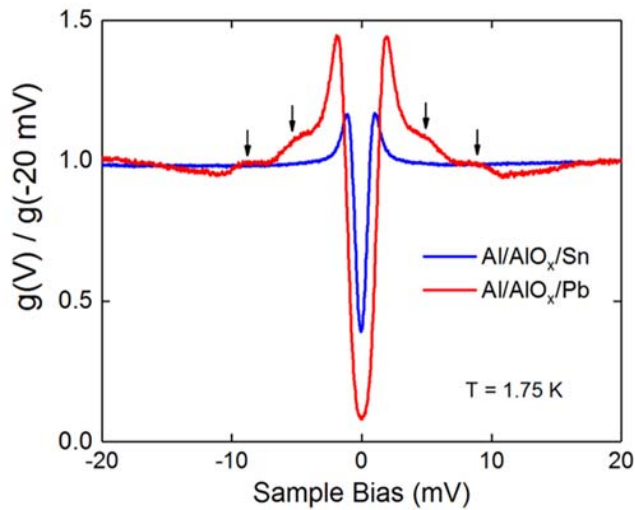

**Figure S6. Comparison of tunneling conductance in two AlO$_x$-based junctions with different counter-electrodes, Pb and Sn.** As indicated by the arrows, the Pb tunnel junction exhibits well-known phonon features above the gap due to the strong electron-phonon coupling underlying the superconductivity in Pb. In contrast, such features are completely missing in the weak-coupled superconductor Sn as the coupling is much weaker.

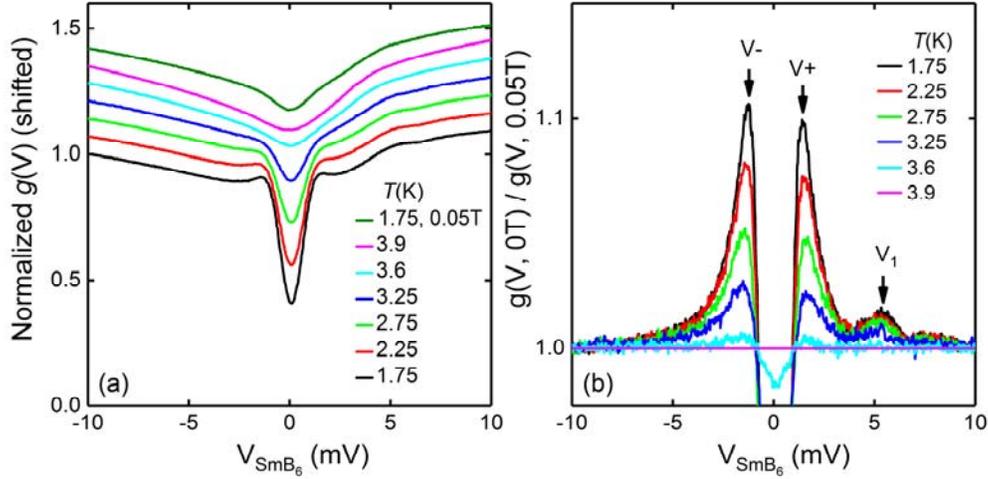

**Figure S7. Evidence for the spin excitons in SmB$_6$/I/Sn tunnel junction.** (a) Temperature dependence of the tunneling conductance with the Sn superconducting (zero field) except for the topmost curve. (b) Temperature dependence of the conductance in (a) normalized by the conductance with the Sn driven normal by a low magnetic field (500 Oe). An additional peak denoted as V$_1$ is still observed and only in the positive-bias branch, similarly to the case of Pb counter-electrode. In addition, the coherence peaks show systematic asymmetry. This counter-electrode independence of the characteristic tunneling features indicates that spin excitons in SmB$_6$ are indeed involved in the inelastic tunneling processes [13].

## References

[1]     D. J. Kim, J. Xia and Z. Fisk, Nat. Mater. **13**, 466 (2014).

[2]     W. A. Phelan, S. M. Koohpayeh, P. Cottingham, J. A. Tutmaher, J. C. Leiner, M. D. Lumsden, C. M. Lavelle, X. P. Wang, C. Hoffmann, M. A. Siegler, N. Haldolaarachchige, D. P. Young and T. M. McQueen, Sci. Rep. **6**, 20860 (2016).

[3]     L. Sun, D. J. Kim, Z. Fisk and W. K. Park, Phys. Rev. B **95**, 195129 (2017).

[4]     W. F. Brinkman, R. C. Dynes and J. M. Rowell, J. Appl. Phys. **41**, (1970).

[5]     D. Li and W. Y. Ching, Phys. Rev. B **54**, 13616 (1996).

[6]     J. Denlinger, J. W. Allen, J.-S. Kang, K. Sun, J.-W. Kim, J. H. Shim, B. I. Min, D.-J. Kim and Z. Fisk, arXiv:1312.6637 (2013).

[7]     E. Frantzeskakis, N. de Jong, B. Zwartsenberg, Y. K. Huang, Y. Pan, X. Zhang, J. X. Zhang, F. X. Zhang, L. H. Bao, O. Tegus, A. Varykhalov, A. de Visser and M. S. Golden, Phys. Rev. X **3**, 041024 (2013).

[8]     J. Jiang, S. Li, T. Zhang, Z. Sun, F. Chen, Z. R. Ye, M. Xu, Q. Q. Ge, S. Y. Tan, X. H. Niu, M. Xia, B. P. Xie, Y. F. Li, X. H. Chen, H. H. Wen and D. L. Feng, Nat. Commun. **4**, 3010 (2013).

[9]     C. H. Min, P. Lutz, S. Fiedler, B. Y. Kang, B. K. Cho, H. D. Kim, H. Bentmann and F. Reinert, Phys. Rev. Lett. **112**, 226402 (2014).

[10]    H. Miyazaki, T. Hajiri, T. Ito, S. Kunii and S. Kimura, Phys. Rev. B **86**, 075105 (2012).

[11]    M. Neupane, N. Alidoust, S. Y. Xu, T. Kondo, Y. Ishida, D. J. Kim, C. Liu, I. Belopolski, Y. J. Jo, T. R. Chang, H. T. Jeng, T. Durakiewicz, L. Balicas, H. Lin, A. Bansil, S. Shin, Z. Fisk and M. Z. Hasan, Nat. Commun. **4**, 2991 (2013).

[12]    N. Xu, P. K. Biswas, J. H. Dil, R. S. Dhaka, G. Landolt, S. Muff, C. E. Matt, X. Shi, N. C. Plumb, M. Radovic, E. Pomjakushina, K. Conder, A. Amato, S. V. Borisenko, R. Yu, H. M. Weng, Z. Fang, X. Dai, J. Mesot, H. Ding and M. Shi, Nat. Commun. **5**, 4566 (2014).

[13]    W. K. Park, L. Sun, A. Noddings, D.-J. Kim, Z. Fisk and L. H. Greene, Proc. Natl. Acad. Sci. U. S. A. **113**, 6599 (2016).

[14]    S. Rössler, T. H. Jang, D. J. Kim, L. H. Tjeng, Z. Fisk, F. Steglich and S. Wirth, Proc. Natl. Acad. Sci. U. S. A. **111**, 4798 (2014).

[15]    V. Madhavan, W. Chen, T. Jamneala, M. F. Crommie and N. S. Wingreen, Science **280**, 567 (1998).

[16]    P. Cheng, C. L. Song, T. Zhang, Y. Y. Zhang, Y. L. Wang, J. F. Jia, J. Wang, Y. Y. Wang, B. F. Zhu, X. Chen, X. C. Ma, K. He, L. L. Wang, X. Dai, Z. Fang, X. C. Xie, X. L. Qi, C. X. Liu, S. C. Zhang and Q. K. Xue, Phys. Rev. Lett. **105**, 076801 (2010).

[17]    T. Hanaguri, K. Igarashi, M. Kawamura, H. Takagi and T. Sasagawa, Phys. Rev. B **82**, 081305 (2010).

[18]    W. Ruan, C. Ye, M. H. Guo, F. Chen, X. H. Chen, G. M. Zhang and Y. Y. Wang, Phys. Rev. Lett. **112**, 136401 (2014).

[19]    Michael M. Yee, Yang He, Anjan Soumyanarayanan, Dae-Jeong Kim, Zachary Fisk and Jennifer E. Hoffman, arXiv:1308.1085 (2013).

[20]    X. H. Zhang, N. P. Butch, P. Syers, S. Ziemak, R. L. Greene and J. Paglione, Phys. Rev. X **3**, 011011 (2013).

[21]    M. Dzero, J. Xia, V. Galitski and P. Coleman, Annu. Rev. Condens. Matter Phys. **7**, 249 (2016).

[22]    U. Fano, Phys. Rev. **124**, 1866 (1961).

[23]    M. Maltseva, M. Dzero and P. Coleman, Phys. Rev. Lett. **103**, 206402 (2009).